# High pressure synthesis of late rare earth *R*FeAs(O,F) superconductors; *R* = Tb and Dy.

**Jan-Willem G. Bos,** [a,b] **George B. S. Penny,** [a,b] **Jennifer A. Rodgers,** [a,b] **Dmitry A. Sokolov,** [a,c] **Andrew D. Huxley** [a,c] **and J. Paul Attfield*** [a,b]

[a] Centre for Science at Extreme Conditions, University of Edinburgh, King's Buildings, Mayfield Road, Edinburgh, EH9 3JZ. Fax: +44 131 651 7049; Tel: +44 131 651 7229; E-mail: j.p.attfield@ed.ac.uk

[b] School of Chemistry, University of Edinburgh, King's Buildings, West Mains Road, Edinburgh, EH9 3JJ.

[c] SUPA, School of Physics, University of Edinburgh, King's Buildings, Mayfield Road, Edinburgh, EH9 3JZ.

**New TbFeAs(O,F) and DyFeAs(O,F) superconductors with critical temperatures $T_c$ = 46 and 45 K and very high critical fields ≥100 T have been prepared at 1100-1150°C and 10-12 GPa, demonstrating that high pressure may be used to synthesisie late rare earth derivatives of the recently reported *R*FeAs(O,F) (*R* = La – Nd, Sm, Gd) high temperature superconductors.**

A breakthrough in high temperature superconductivity has recently occurred with the discovery that rare earth oxypnictides *R*FeAsO (first reported for *R* = La, Ce, Pr, Nd, Sm and Gd)[1] can show critical temperatures surpassed only by the high-$T_c$ cuprates. These materials have a tetragonal, layered crystal structure as depicted in the inset to Fig. 1. Superconductivity has been induced by the partial substitution of fluoride into the *R*O layers, which leads to electron doping (reduction of iron) in the electronically active FeAs slabs. The first report of superconductivity was in LaFeAsO$_{1-x}$F$_x$ samples with $T_c$'s up to 26 K,[2] increasing to 43 K at 4 GPa pressure.[3] Superconductivity has subsequently been induced in the other members of the *R*FeAsO series using fluoride doping, with ambient pressure $T_c$'s of 41 K for *R* = Ce,[4] 52 K for Pr[5] and Nd,[6] 43-55 K for Sm samples,[7] and 36 K for Gd.[8]

High pressure and temperature synthesis is known to stabilise many late rare earth analogues of early rare earth solid compounds. This approach has been used to explore the stabilisation of TbFeAs(O,F) and DyFeAs(O,F) phases. Polycrystalline samples of nominal compositions $R$FeAsO$_{1-x}$F$_x$ ($R$ = Tb, Dy; x = 0, 0.1, 0.2) were synthesised from stoichiometric amounts of $R$As, Fe$_2$O$_3$, FeF$_2$ and Fe. TbAs and DyAs were prepared from a stoichiometric mixture of the elements heated to 500 °C for 5 hours and then 900 °C for 10 hours in an evacuated quartz tube. All chemicals were obtained from Sigma Aldrich with at least 99.9% purity. The reactants were mixed and ground in a glove box, sealed in a BN capsule, and subjected to pressures of 10 GPa ($R$ = Tb) or 12 GPa ($R$ = Dy) using a Walker two-stage multianvil within a 1000 tonne press. Once at pressure, the samples were heated to 1100-1150 °C in 10 min, held at this temperature for 20 min, and then quenched to room temperature, followed by release of the pressure. The products were dense, black, sintered polycrystalline pellets and were characterised by powder X-ray diffraction, (Fig. 1) magnetisation (Fig. 2) and resistivity (Fig. 3) measurements.[‡]

The $R$ = Tb samples all contained the tetragonal $R$FeAsO type phase with traces of TbAs (Fig. 1). The synthesis of DyFeAsO was unsuccessful but DyFeAsO$_{1-x}$F$_x$ phases were obtained for x = 0.1, and 0.2 with DyAs also present (see Supplementary Figures). All four fluoride-doped samples show both magnetic and resistive superconducting transitions, with critical temperatures of 40-46 K. Fig. 2 shows that the samples are bulk superconductors, with some reduction from the theoretical full diamagnetism due to the presence of impurities and field penetration into small grains. The refined lattice parameters and $T_c$ values are shown in Table 1. We also synthesised a new TbFeAsO$_{0.9}$ analogue of the reported oxygen-deficient $R$FeAsO$_{0.85}$ superconductors at 10 GPa.[9] This sample is superconducting with $T_c$ = 50 K, further details will be reported elsewhere.

The resistivities show clear transitions to zero resistance (Fig. 3) with a smooth negative curvature of the resistivity in the normal state. This differs from data for other superconducting oxypnictides that appear to show higher temperature transitions.[10] Changes in this behaviour are theoretically predicted to be very sensitive to competing energy scales controlling the physics of these materials.[11] The resistive transition width increases with magnetic field for all samples as observed in other oxypnictides,[12] consistent with a large anisotropy of the critical field, reflecting the structural and electronic anisotropy. The upper critical field $B_{c2}$ increases to 9 T in <2 K below $T_c$ for TbFeAsO$_{0.8}$F$_{0.2}$ (Fig. 3 upper inset) and, in BCS theory neglecting paramagnetic limitation, this corresponds to $B_{c2}$ exceeding 100 T at low temperatures. Taking the onset of the transition to give the upper critical field for superconductivity in the most favourable direction (parallel to the FeAs planes) an upper estimate for the superconducting coherence length perpendicular to this direction is 13(1) Å. This corresponds to the geometric mean of the in-plane and out-of-plane coherence lengths. Given that the anisotropy is large, the out-of-plane value is therefore likely to be significantly smaller than the FeAs layer spacing, demonstrating that superconductivity is strongly 2-dimensional. The zero resistance transition field has a noticeably more marked upward curvature at low field than observed for the transition onset. This might reflect a transition to a vortex liquid state, which is well known in the high-$T_c$ copper oxide superconductors, or be an indication of multiple band superconductivity as established in $MgB_2$.[13]

The $T_c$'s of the $R$FeAsO$_{1-x}$F$_x$ ($R$ = Tb, Dy, x = 0.1, 0.2) samples do not differ greatly and there is no clear trend in the lattice parameters with x, showing that the actual range of doping may be more limited that in the nominal compositions. Further work will be needed to determine the precise range of x and optimise phase purity. It is notable that the $T_c$'s of TbFeAs(O,F) and DyFeAs(O,F) are comparable to those of the early $R$ = Ce, Pr, Nd, Sm materials. The lower value of 36 K reported for GdFeAs(O,F)[8] suggested that superconductivity might be suppressed as the rare earth size decreases, but the present results show that the superconducting properties change little between Ce and Dy. It will be important to explore further $R$FeAs(O,F) superconductors of the heavy rare earths to discover how superconductivity develops across the entire series.

‡     Powder X-ray diffraction data were collected on a Bruker AXS D8 diffractometer using Cu K$\alpha_1$ radiation. Data were recorded at 10 ≤ 2θ ≤ 100° with a step size of 0.007° for Rietveld analysis. The ac magnetic susceptibility was measured from 3 to 50 K with a field of 0.5 Oe oscillating at 117 Hz using a Quantum Design superconducting quantum interference device magnetometer. The electrical resistivity was measured by the conventional four-probe method between 1.7 and 300 K using a Quantum Design physical property measurement system.

**Table 1.** Cell parameters and volume and $T_c$'s from the onset of diamagnetism and the resistive transition mid-point for $R$FeAs(O,F).

| $R$FeAs(O,F) | $a$/Å | $c$/Å | Vol./Å$^3$ | $T_C(\chi_{ons})$/K | $T_c(\rho_{mid})$/K |
|---|---|---|---|---|---|
| TbFeAsO | 3.8632(8) | 8.322(3) | 124.20(8) | - | - |
| TbFeAsO$_{0.9}$F$_{0.1}$ | 3.8634(3) | 8.333(1) | 124.38(3) | 45.5 | 43.8 |
| TbFeAsO$_{0.8}$F$_{0.2}$ | 3.860(2) | 8.332(6) | 124.2(2) | 45.2 | 45.9 |
| DyFeAsO$_{0.9}$F$_{0.1}$ | 3.8425(3) | 8.2837(8) | 122.30(3) | 45.3 | 45.4 |
| DyFeAsO$_{0.8}$F$_{0.2}$ | 3.8530(3) | 8.299(1) | 123.21(2) | 43.0 | 43.0 |

**Fig. 1**. Rietveld fit to the x-ray diffraction profile of TbFeAsO$_{0.9}$F$_{0.1}$, with Bragg reflection markers shown below those for the minority phase TbAs. An additional impurity peak is observed at 38° 2θ. Refinement residuals are $R_{wp}$ = 2.63%, $R_p$ = 2.00% and $\chi^2$ = 1.64 for 24 variables. Atom positions (x,y,z) and isotropic-U's; Tb (¼,¼,0.1447(4)), 0.003(1) Å$^2$; As (¼,¼,0.6654(6)), 0.009(2) Å$^2$; Fe (¾,¼,½), 0.003(1) Å$^2$; O,F (¾,¼,0), 0.07(1) Å$^2$. The inset shows the structure.

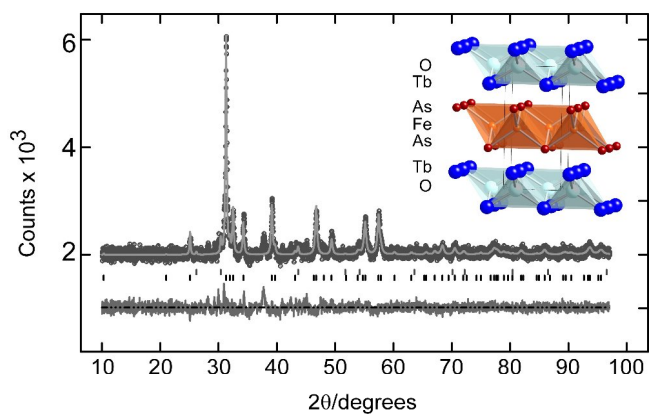

**Fig. 2** Ac magnetic volume susceptibility vs. temperature for $R$FeAs(O,F) ($R$ = Tb, Dy) samples; χ' = -1 corresponds to full diamagnetism.

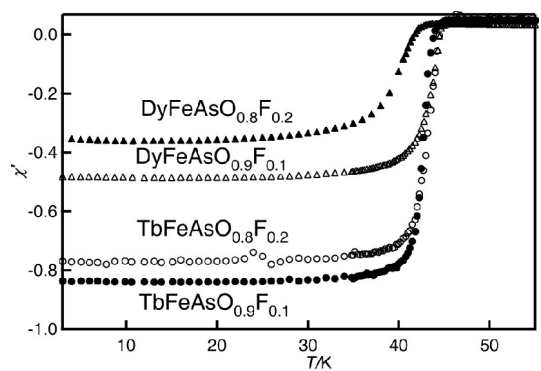

**Fig. 3** Temperature dependence of the resistivity of TbFeAsO$_{0.8}$F$_{0.2}$. The lower inset shows the superconducting transitions in zero and 9 T fields, and the upper inset shows the onset (0 %) of the transition and the zero resistance point (100%) indicative of the upper critical field (B$_{c2}$). The coherence length is obtained from the fit to the latter values.

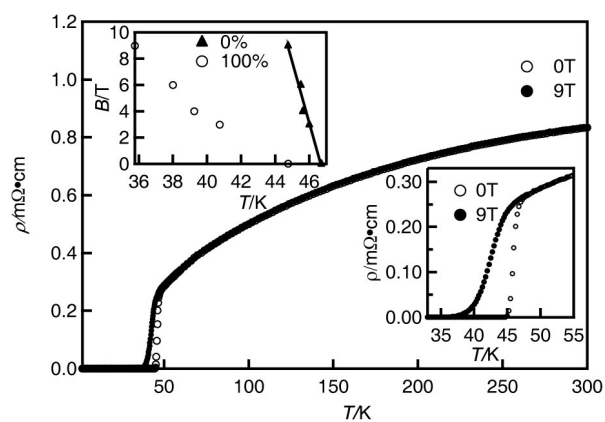

**Supplementary Figure 1** Observed (•) and calculated (−) x-ray diffraction profiles for DyFeAsO$_{0.9}$F$_{0.1}$ at room temperature. The upper set of tick marks refers to the minority phase, DyAs and the lower set to DyFeAsO$_{0.9}$F$_{0.1}$. Goodness of fit parameters of $\chi^2$ = 1.76, $R_{wp}$ = 2.33%, $R_p$ = 1.75%, for refinement of 24 variables.

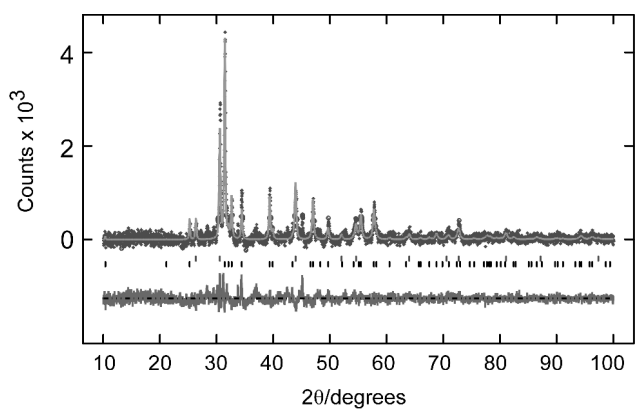

**Supplementary Figure 2** Temperature dependence of the electrical resistivity of TbFeAs(O,F) and DyFeAs(O,F) measured in zero field. Inset: Expanded view of low temperatures, showing superconducting transitions.

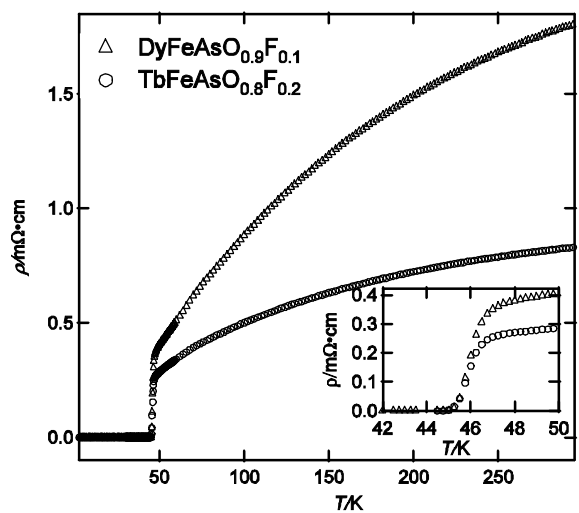